\documentclass{aa}  

\usepackage{graphicx}
\usepackage{gensymb}
\usepackage{amsmath}
\usepackage{caption}
\usepackage{subcaption}

\usepackage{txfonts}
\usepackage[hidelinks]{hyperref}
\hypersetup{colorlinks=true,allcolors=cyan}

\begin{document}

   \title{Radio properties of the quasi-periodic eruption source RXJ1301.9+2747 at parsec scales}
   \author{S.~D. von Fellenberg\thanks{Feodor Lynen Fellow of the Alexander von Humboldt Foundation}\inst{1,2}
   \and R. Arcodia\thanks{NASA Einstein fellow}\inst{3}
   \and P. Benke\inst{2}
   \and A. Goodwin\inst{4}
   \and Y. Y. Kovalev\inst{2}
   \and E. Ros\inst{2}
   \and M. Janssen\inst{5}
   \and M. Giustini\inst{6}
   \and G. Miniutti\inst{6}
}

   \institute{
    Canadian Institute for Theoretical Astrophysics, University of Toronto, 60 St. George Street, Toronto, ON M5S3H8, Canada
    \and
   Max-Planck-Institut f\"ur Radioastronomie, Auf dem H\"ugel 69, D-53121 Bonn, Germany
   \and 
   Kavli Institute for Astrophysics and Space Research, Massachusetts Institute of Technology, Cambridge, MA 02139, USA
   \and
   International Centre for Radio Astronomy Research – Curtin University, GPO Box U1987, Perth, WA 6845, Australia
   \and
    Department of Astrophysics, Institute for Mathematics, Astrophysics and Particle Physics (IMAPP), Radboud University, P.O. Box 9010, 6500 GL Nijmegen, The Netherlands 
   \and
   Centro de Astrobiolog\'ia (CAB), CSIC-INTA, Camino Bajo del Castillo s/n, 28692 Villanueva de la Ca\~nada, Madrid, Spain
   }

   \date{Received September \today; accepted \today}

  \abstract
   {\\\\Quasi-periodic eruptions (QPEs) are repeating soft X-ray flares associated with galactic nuclei. Several recent works have found evidence that the accretion flow in the galactic nuclei of QPEs is of recent origin, and that it is unlike canonical active galactic nuclei (AGN). A precursor tidal disruption event has been observed in a few cases. In this work we report new radio observations of the QPE host galaxy RXJ\,1301.9+2747 taken at 5.0\,GHz with the High Sensitivity Array (HSA), to complement archival 1.7\,GHz observations reported previously. Our new observations confirm the presence of a highly compact radio source in RXJ\,1301.9+2747, which is smaller than $0.9\times0.4~\mathrm{pc}$ at 5.0\,GHz. {The nonsimultaneous very long baseline interferometry (VLBI) compact flux of the source is consistent {with} a negative spectral index, and thus is similar to the larger non-VLBI scale radio spectral index.}  {Contrary to earlier results at 1.7\,GHz, we find the 5\,GHz emission offset from the optical Gaia position, which may be due to dust extinction in the host galaxy.} In addition, there is a significant offset between the 1.7 and 5.0\,GHz data, which may result from astrophysical uncertainties in the calibration source. This sheds new light on the elusive properties of the radio-detected QPE sources. Consistent with previous results, our observations disfavor a star formation or jet-core-region origin of the radio emission. {We cannot rule out a reconnection-driven scenario for the radio emission, but} we favor a remnant jet or outflow scenario. This is overall in agreement with the radio properties of radio-detected QPE sources at lower angular resolution.}
   \keywords{}
   \maketitle

\section{Introduction}

Quasi-periodic eruptions (QPEs) are high-amplitude soft X-ray flares, to date found in the nuclei of about a dozen galaxies \citep{Miniutti+2019:qpe1,Giustini+2020:qpe2,Arcodia+2021:eroqpes,Arcodia+2024:eroqpes,Chakraborty+2021:qpe5cand,Chakraborty+2025:upj,Quintin+2023:tormund,Bykov+2024:tormund,Nicholl+2024:qiz,Hernandez-Garcia+2025:ansky,Arcodia+2025:ero5}. QPE flares evolve on  timescales of hours to days, with a duty cycle (i.e., the ratio of their duration to recurrence {time) of} $\approx10-20\%$ \citep[e.g.,][]{Arcodia+2024:eroqpes,Chakraborty+2025:upj,Nicholl+2024:qiz,Hernandez-Garcia+2025:ansky,Arcodia+2025:ero5}. The eruptions are characterized by a soft quasi-thermal spectral shape that appears harder during the rise compared to the decay at a compatible luminosity, which implies an increase in the size of the emitting region during the flares assuming a simplistic blackbody emission model \citep{Arcodia+2022:ero1,Miniutti+2023:gsn,Arcodia+2024:eroqpes,Chakraborty+2025:upj,Nicholl+2024:qiz,Hernandez-Garcia+2025:ansky}. To date, no counterpart to the X-ray flares has been found at other wavelengths, including optical, UV, IR, and radio \citep[e.g.,][for some of the latest attempts at higher angular resolution]{Giustini+2024:rxj,Nicholl+2024:qiz,Wevers+2025:sed}. In quiescence (i.e., when not in eruption), these nuclei are normally detected in the soft X-rays, as well as in the UV and optical bands, which points to the presence of a radiatively efficient accretion disk \citep{Nicholl+2024:qiz,Wevers+2025:sed,Guolo+2025:sed,Guolo+2025:gsnlongterm}. This disk is found to be quite compact in size and that, at least in some cases, it has originated from a previous tidal disruption event (TDE) in the nucleus \citep{Quintin+2023:tormund,Miniutti+2023:gsn,Bykov+2024:tormund,Nicholl+2024:qiz,Chakraborty+2025:upj,Guolo+2025:gsnlongterm}.

Significant attention is being dedicated to QPE models that involve an extreme mass-ratio in-spiral (EMRI) system, in which the flares are produced when the stellar-mass secondary pierces through the accretion disk around the primary supermassive black hole \citep[SMBH;][]{Xian+2021,Linial+2023:tdeemri,Franchini+2023,Tagawa+2023,Zhou+2024:qpemodel,Zhou+2024:longterm,Vurm+2024:emission,Huang+2025:simul}. 
Within the framework of this popular secondary-disk collision model, QPE emission is dominated by soft X-rays with contributions down to the optical and UV (likely overwhelmed by disk emission and stellar contamination) with little, if any, associated emission at radio wavelengths \citep{Vurm+2024:emission}.
The discovery QPE source GSN\,069 did not show any correlated variability in the radio \citep{Miniutti+2019:qpe1}, and recently \citet{Giustini+2024:rxj} reported the lack of any radio variability associated with the QPE flares with Very Large Array (VLA) data covering five QPEs of the source RXJ\,1301.9+2747, confirming the absence of a low-frequency tail of the QPE emission. This is in agreement with \citet{Goodwin+2025:radioqpe}, who reported a more systematic radio campaign of most of the known QPE sources, although with data that were not strictly simultaneous. Thus, it is our current understanding that any non-X-ray observation of QPE sources traces either the underlying nuclear emission from the SMBH's accretion flow, or the emission from the host galaxy, depending on the angular resolution and sensitivity used. However, we note that a canonical X-ray corona (often seen as ubiquitous in accreting black holes) has not been significantly detected in any QPE source yet, even though weak residuals to just the accretion disk have been reported before \citep[e.g.,][]{Miniutti+2019:qpe1,Miniutti+2023:gsn,Giustini+2024:rxj,Arcodia+2024:eroqpes}. Furthermore, the recent systematic study of \citet{Goodwin+2025:radioqpe} shows that the properties of five radio-detected QPE sources, which are not related to the eruptions themselves, are generally inconsistent with canonical radio AGN emission, but may be tantalizingly interpreted as remnant TDE-like outflows. Together with the absence of a broad-line region and a torus evident since the first discoveries of QPE sources \citep{Shu2017,Miniutti+2019:qpe1,Giustini+2020:qpe2,Arcodia+2021:eroqpes,Wevers+2022:hosts}, this all points toward a relatively unusual accretion system compared to typical AGN. 

For this work  we took a closer look at the nuclear radio properties of the QPE source RXJ\,1301.9+2747 (hereafter RXJ) with very long baseline interferometry (VLBI) observations. RXJ was the second source where QPE were identified as such \citep{Giustini+2020:qpe2}, with archival eruptions being detected years \citep{Sun+2013:rxj,Shu2017} and possibly decades earlier \citep{Dewangan+2000:rxj}. It is highly likely that its quiescent soft X-ray radiatively efficient accretion disk dates back to at least the 1980s as a soft source was detected by EXOSAT \citep{Branduardi-Raymont+1985:exosat,Giustini+2024:rxj}. It is also the best studied QPE source in the radio band \citep[e.g.,][]{Yang2022}, showing variability over a timescale of days and a steep (spectral index $\sim-0.78$) time-averaged spectrum between 0.89 and 14\,GHz. \citet{Yang2022} obtained the first 1.7\,GHz VLBA observations of the source and found that the source is compact, and argued against a star formation favoring episodic jet ejections driven by magnetic reconnection.

\section{Data}

The QPE source RXJ was first observed with VLBI on February 14, 2017, using the Very Long Baseline Array (VLBA, program ID BS255 \citet{Yang2022}). 
The observations were carried out for 12 hours and included the entire VLBA. The data were phase-referenced using FBQS J1300+2830, with $180~\mathrm{s}$ scans on RXJ and $60~\mathrm{s}$ scans on J1300+2830. The overall data quality is good across the array. 
However, several stations were affected by strong radio frequency interference (RFI) and had to be flagged.
The VLBI data were processed using the Astronomical Image Processing System (AIPS) \citep{GreisenAips2003} software; and we performed consistency checks with \textit{rPicard} \citep{Janssen2019A}. To obtain a light curve from the phased-VLA observations the VLA data was reduced with CASA \citep{CASA_McMullin2007}.

We obtained new VLBI observation of RXJ on February 5, 2023, using the High Sensitivity Array (HSA; program ID BV087). 
The observations were carried out for 12 hours and included the entire VLBA, as well as the phased-VLA and the Effelsberg observatory. The data were phase-referenced using FBQS J1300+2830, with $180\,$s scans on RXJ and $60\,$s scans on J1300+2830. 
The overall HSA-data quality is very good across the array. However, the phased-VLA were affected by poor phasing for the first 10 seconds of each calibrator scan, which we flagged. 
We processed the data with AIPS, and we used \textsc{rPicard} to check for consistency.

Finally, we obtained an additional simultaneous set of VLBA observations on February 16, 2024 (program ID BV092). The observations covered three bands, L, C, and X, in a single track, which is why they have a low signal-to-noise ratio (S/N). The source could be detected in 1.7\,GHz L and 5.0\,GHz C band, but not in the 8.0\,GHz X band. The detections in both bands are marginal, and the source is only discernible if a uniform weighting \citep[e.g.,][]{Mueller2023} scheme is chosen. 
As before, the data were calibrated in AIPS and \textsc{rPicard}. An overview of the VLBI observations used in this paper is given in \autoref{tab:obs}.

\begin{table*}[]
    \centering
    \caption{Overview of observations used in this paper.}
    \begin{tabular}{c|c|c|c|c|}
         Date & Frequency & Array & Reference & Detection \\
         \hline
         14 Feb. 2017 & 1.7\,GHz & VLBA & \citep{Yang2022} & yes \\
         5 Feb. 2023 & 5.0\,GHz & HSA & this paper & yes \\
         16 Feb. 2024 & 1.7\,GHz & VLBA & this paper & partial\\
         16 Feb. 2024 & 5.0\,GHz & VLBA & this paper & partial\\
         16 Feb. 2024 & 8.0\,GHz & VLBA & this paper & no\\
         
    \end{tabular}
    \label{tab:obs}
\end{table*}

\section{Results}
\subsection{Parsec-scale emission at 5.0\,GHz}
\begin{figure}
    \centering
    \includegraphics[width=0.485\textwidth]{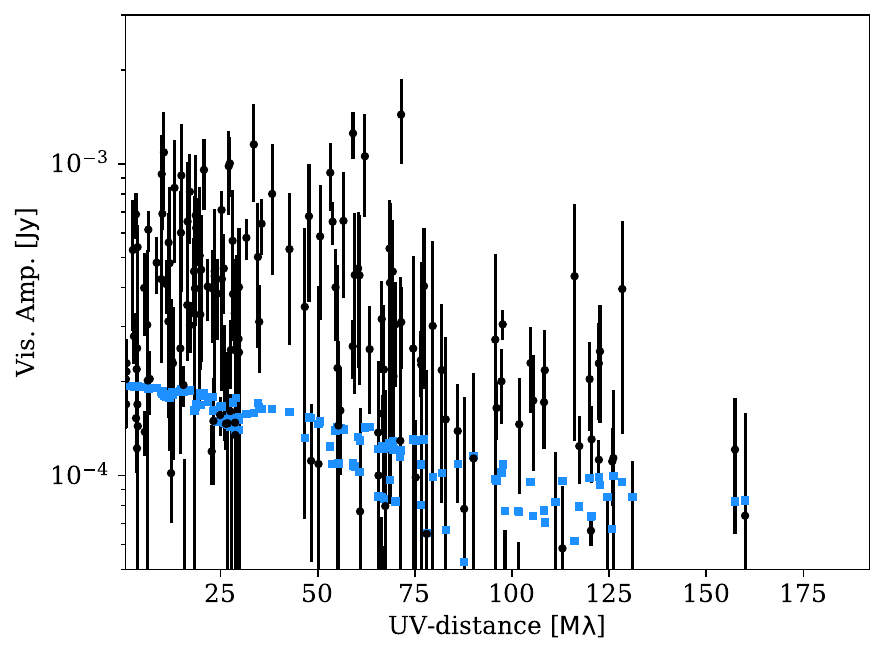}
    \caption{Time averaged visibilities as observed by the HSA in February 2023 at 5\,GHz. The blue points illustrate the best fit \textsc{ehtim} image model, shown in \autoref{fig:clean_map}.}
    \label{fig:visbilities}
\end{figure}

\begin{figure}
    \centering
    \includegraphics[width=0.485\textwidth]{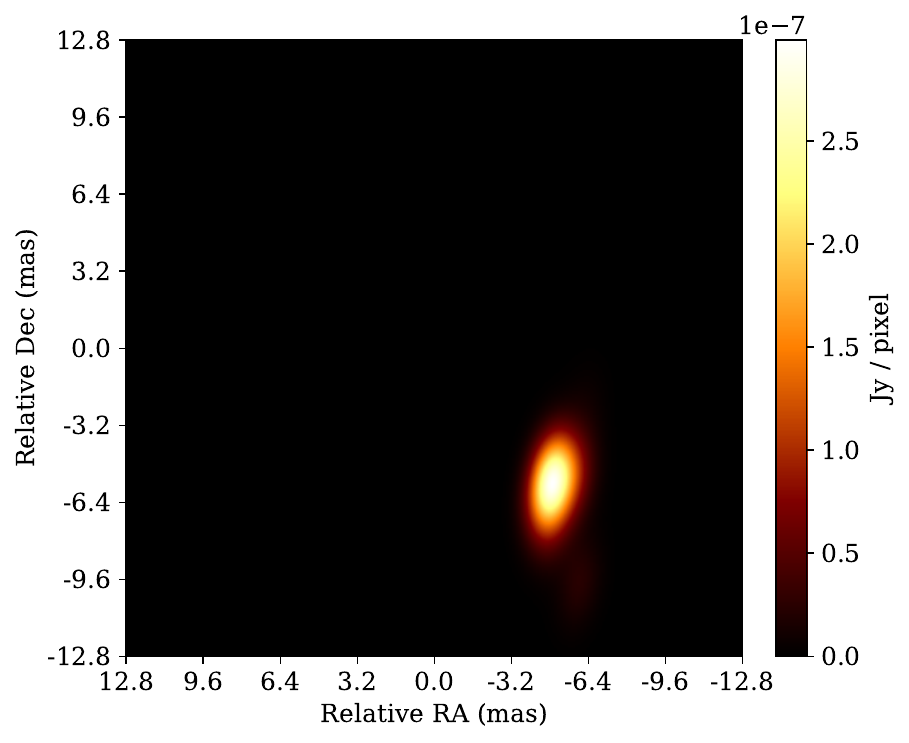}
    \caption{Map of RXJ obtained with \textsc{ehtim}, showing the 5.0\,GHz HSA observations. The data are offset to the observation phase center at R.A.$=13~02~ 00.138$, Dec.$=+27~46 ~57.855$. The restoring beam sized use in the \textsc{ehtim} image is $(2.9\times1.4)~\mathrm{mas}^2$.}
    \label{fig:clean_map}
\end{figure}

In order to derive new constraints on the compact emission, and leverage the much higher spatial resolution in the HSA data, we used the standard CLEAN implementation in Difmap \citep{Shepherd1997} to model and image the data. In addition, we used the RML imaging code \textsc{ehtim} to image the data \citep[][]{Chael2018}. The visibilities are shown in \autoref{fig:visbilities}; an \textsc{ehtim} image is shown in \autoref{fig:clean_map}, which shows a source detected with a $S/N> 100$. While the self-calibrated visibility amplitude shows a small drop in flux, potentially indicating marginally resolved emission, we caution that even small gain calibration errors can lead to such a visibility signature. Thus, we conservatively treated the restoring beam size of the clean image (\autoref{fig:clean_map_difmap}, $\theta_{\rm{beam}} = (1.7\times 0.8)~\mathrm{mas}$) in the self-calibrated images as an upper limit of spatial extent of the source. 

Similarly, any flux density values can be biased by the amplitude self-calibration. Therefore, we derived the flux density constraint using data that are not self-calibrated. We fit the data with a circular Gaussian component. The flux density of the component is estimated to be $180\pm5~\mathrm{\mu Jy}$, where the uncertainty is given by the statistical error of the fit. The size of the component is estimated at $(0.48\pm0.3)~\mathrm{mas}$; the fit $\chi^2{\rm{red.}}$ is  $\chi^2_{\rm{red.}}=1.6$.

Using the more conservative size limit, the restoring beam size, we obtained lower limits on the brightness temperature given by
\begin{equation}
    \rm{T_{B}}=1.22\times 10^{12}\dfrac{F_{\rm{5.0~\mathrm{GHz}}}}{\nu^2 \theta_{\rm{maj}}\theta_{\rm{min}}} ~\mathrm{[K]} > 2.2\times 10^{7}~\mathrm{K}.
\end{equation}

\subsection{Compact flux and spectrum of RXJ}
\cite{Yang2022} report a negative spectral measurement of $F_{\nu=\rm{radio}} \propto \nu^{-0.78\pm0.03}$, which is based on non-simultaneous GMRT, VLA, and VLBA measurements between 0.89\,GHz and 14\,GHz. At the same time, \cite{Yang2022} report that the source shows significant variability, which implies that the spectral measurement may be biased by the variability. \cite{Goodwin+2025:radioqpe}, however, confirm the negative spectral index by (quasi)-simultaneous observations, which show a consistent negative spectral index $F_{\nu=\rm{radio}} \propto \nu^{-0.684\pm0.003}$. 
Large-scale VLA observations still do not directly probe the compact emission probed at VLBI scales, and therefore larger scale emission from a jet or outflow may bias the VLA-spectral index measurement. Here we provide three new lines of evidence that there is  an overall negative spectral index, and that the compact flux dominates the emission seen by the VLA and other instruments.

First, we  identified RXJ in the LoFAR DR2 catalog \citep{Shimwell2022}, mosaic \textit{P195+27}. The source has $143~\mathrm{MHz}$ flux density of $(1.1\pm0.1)~\mathrm{mJy}$, which is clearly higher. However, because the LoFAR beam is substantially larger, these measurements have little predictive power for the compact scales observed with VLBI.

Second, we measured a flux density of $F_{\nu=5.0~\rm{GHz}}=180\pm 5~\mathrm{\mu Jy}$ from the 5\,GHz HSA observations. This value is consistent with the flux density  measured simultaneously by the phased-VLA data\footnote{Phased-VLA observations are carried out as part of the HSA-observations; the connected-element data was reduced to obtain a flux measurement.} ($F_{\rm{p-VLA}}=100\pm80~\mathrm{\mu Jy}$). This shows that, within the errors, all observed radio flux stems from the compact emission, and thus the VLA-derived spectral index is applicable to the compact emission as well.

Third, we obtained simultaneous VLBA observations of RXJ in February 2024. Unfortunately, the sensitivity of these observations is limited, and thus the detections are marginal and depend on the choice of error weighting scheme. 
Under these assumptions, the peak flux is $(41\pm9)~\mathrm{\mu Jy/beam}$ in the L-band map, and $(10\pm4)~\mathrm{\mu Jy/beam}$ in the C-band map. The noise level in all three bands is similar ($\sim 4~\mathrm{\mu Jy}$), which again confirms the negative spectral index, but because of the low sensitivity of the data the VLBA results are not robust. Nevertheless, all observations indicate that the results derived by \cite{Yang2022} and \cite{Goodwin+2025:radioqpe} apply to the compact emission, and that  the source indeed exhibits a negative spectral index of $F_\nu \propto \nu^{\sim -0.8}$

\subsection{Astrometry of RXJ}\label{sec:astrometry}
Very long baseline interferometry allows the measurement of the relative position of the source with high astrometric accuracy. Because astrometry can be carried out at a much greater accuracy than the width of the beam, it allows us to measure the evolution of the source over time and frequency. One caveat of VLBI astrometric observations is that they measure the on-sky separation between two sources. 
This on-sky separation is strictly the separation between the center of light of the two sources at a given frequency, which may not directly relate to physically meaningful quantities such as the location of the SMBH.

In our case, we measure the distance between the center of light of RXJ and the calibrator source J1300+2830. Therefore, when comparing the RXJ position as a function of frequency, any frequency-dependent shift of the apparent on-sky location of the calibrator will affect the measured position.
J1300+2830 is a compact AGN that shows a jet-extension in the northeastern direction (see \autoref{fig:difmap_calib}). Thus, the observed center of light J1300+2830 corresponds to what is typically referred to as the core. Physically, this core position corresponds to the base of the jet, upstream of the actual SMBH location.

The position of this core with respect to the black hole location depends on frequency, where lower frequencies are farther away from the black hole. This is known as the core shift effect \citep[e.g.,][]{Lobanov1998}.
The core shift is caused by a change in optical depth of the expanding jet of the calibrator, which causes different frequencies to appear at different  on-sky locations \citep[e.g.,][]{Blandford1979}.

Thus, in order to measure the position of RXJ as function of frequency, we have to account for the calibrator core shift effect. Specifically, we have to obtain the 1.7 to 5\,GHz calibrator  core shift and subtract it from the apparent sky location. As the effect occurs along the jet axis of the calibrator, we have to subtract it along this direction.

Below, we outline our procedure.\footnote{We provide additional explanatory figures in \autoref{app:coreshift}.} The methodology applied follows the standard procedure applied in astrometric VLBI observations \citep[e.g.,][]{Kovalev2008, Plavin2019_gaiavlbi, Petrov2019_gaia, Benke2024_tan}.

First, we estimated the calibrator core shift by measuring the distance between the core and a bright jet feature at multiple frequencies. To accomplish this, we used the \texttt{astrogeo}\footnote{\url{astrogeo.org}} VLBI Calibrator Archive to obtain maps of the calibrator at 8\,GHz. The jet feature exhibits a negative spectral index, consistent with optically thin emission; therefore, its position should be independent of frequency. Consequently, we can adopt the jet feature as an astrometric reference point to measure the core–jet feature separation across different frequencies.

However, jet features typically exhibit motions of a few milliarcseconds per year \citep[e.g.,][]{Zensus1987}. We confirmed this behavior by analyzing all available archival 8\,GHz data. The \texttt{astrogeo} dataset reveals an approximately linear trajectory of the jet feature, as illustrated in \autoref{fig:jet-knot}.

Thus, to derive the calibrator core shift between 1.7 and 5.0\,GHz, we first measured the separation between the core and the jet feature at 5.0\,GHz in 2023 and at 8.0\,GHz in 2018. We then accounted for the offset introduced by the intrinsic motion of the 8.0\,GHz jet feature between 2018 and 2013. Using these values, we computed the core shift between 5.0 and 8.0\,GHz, and finally, we linearly extrapolated this result to the 1.7\,GHz frequency band, as illustrated in \autoref{fig:core-shift-calibrator}.

In all of these steps, we accumulated measurement errors, which we propagated through numerical Monte Carlo simulation. We obtained a calibrator core shift of $\Delta r_{\rm{calib, 5.0-1.7GHz}} = (0.49\pm0.21)~\mathrm{mas}$.

In addition to these measurement uncertainties, the core shift determination is subject to astrophysical variability, as typical AGN core shifts are not constant and may vary over time. For example, \cite{Plavin2019} reported a median core shift variability of $0.3~\mathrm{mas}$, with the largest deviations occurring primarily during source flaring episodes. The source J1300+2830 did not exhibit flaring in any of the \texttt{astrogeo} observations (core flux densities $f_{\mathrm{core}} \in [0.151, 0.156, 0.201, 0.212, 0.232]~\mathrm{Jy}$), nor were any new jet components ejected. We therefore assume that the systematic uncertainty due to core shift variability does not exceed that associated with extrapolating the calibrator core shift from 5.0\,GHz to 1.7\,GHz.

\begin{figure}
    \centering
    \includegraphics[width=0.485\textwidth]{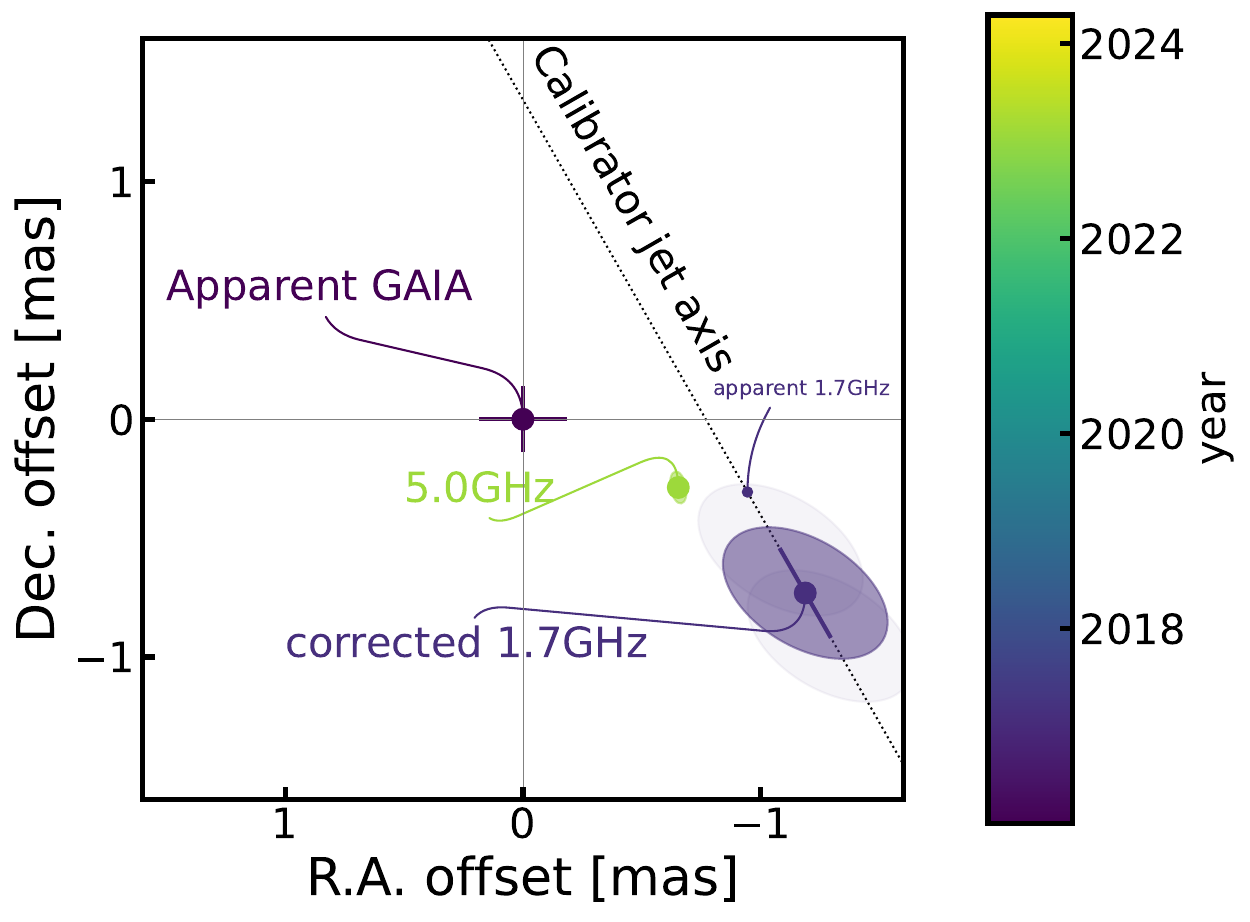}
    \caption{Astrometry of Gaia, and the 1.7\,GHz and 5.0\,GHz observations, as annotated. The Gaia optical position is placed at (0,0); all other positions are plotted relative to this position. The ellipses indicates the best fit Gaussian ellipse derived from a model fit to the VLBI data. Uncertainties are at $1\sigma$ level. The dashed line indicates the calibrator core shift axis along the apparent 1.7GHz position (violet dot) is shifted; the translucent gray ellipses indicate the systematic uncertainty of that core shift (see text for details).}
    \label{fig:astrometry}
\end{figure}

Keeping in mind the caveats mentioned above, we were able to correct the calibrator's core shift and to measure the relative astrometric position of the two emission components relative to each other. \autoref{fig:astrometry} illustrates the relative position of the 1.7 and 5.0\,GHz emission. The emission is relative to the apparent Gaia position. The 1.7 and 5.0\,GHz emissions are significantly offset from one another by $c_{\rm{RXJ}} = (0.69\pm0.49)~\mathrm{mas}$. 

Absent higher frequency data, we cannot estimate the location of the SMBH with respect to the calibrator core at 8\,GHz, which would be needed to obtain an absolute optical to radio reference system. As the calibrator core shift increases the separation between the GAIA and the radio measurements (see \autoref{fig:astrometry}), we determine an upper limit for the relative position of the radio and the Gaia position. Even so,  the 5.0\,GHz and the 1.7\,GHz positions are both significantly offset from the GAIA position.

\section{Discussion}
Our analysis of VLBI observations in the centimeter-radio regime of RXJ has established the following observational aspects:

\begin{itemize}
    \item RXJ emission is extremely compact. Our highest resolution image at 5.0\,GHz implies emission geometry that is smaller than $(1.8\times0.7)~\mathrm{mas}$, corresponding to a physical size of $(0.9\times0.4)~\mathrm{pc}$.
    \item Because we do not resolve the source, we obtain lower limits on the brightness temperatures. The most constraining limit was obtained with the 5.0\,GHz HSA observations, which is $T_{\rm{B}} > 2.2\times 10^7~\mathrm{K}$.
    \item The source shows a parsec-scale astrometric shift between the 2017 1.7\,GHz VLBA and 2023 5.0\,GHz HSA observations (\autoref{fig:astrometry}). The shift is significant at low significance ($1.4\sigma$), although it is uncertain due to astrophysical biases in the calibrator phase reference source. 
    \item The VLBI flux measurements confirm the negative spectral index, but we cannot rule out biases due to variability across the epochs. At the same time, the consistency between the VLA and VLBI flux densities indicates that the compact emission accounts for all of the 5.0\,GHz radio emission of RXJ.
\end{itemize}

This allows us to test three hypotheses for the nature of the emission:

\begin{enumerate}
    \item Star formation scenario: The observed emission is caused by star formation activity (Bremsstrahlung).
    \item LLAGN-core scenario: The emission stems from the core region of a typical low-luminosity AGN (LLAGN), i.e., the nozzle of a parsec-scale jet.
    \item Horizon-scale reconnection: This scenario was proposed by \cite{Yang2022} in which magnetic reconnection at black hole horizon scales produces the observed variability and compact emission.
    \item Remnant outflow or jet scenario: The emission stems from a region of an outflow or jet, with an extinguished or much fainter core. 
\end{enumerate}

\subsection{Star formation scenario}
This scenario is consistent with the negative spectral index of the source; however, as discussed in \cite{Yang2022}, it is difficult to reconcile with the compactness and the resulting high brightness temperature of the source. Star formation processes typically produces brightness temperatures below $10^6~\mathrm{K}$ \citep{PerrezTorres2021}, clearly below the VLBI brightness temperature estimate for RXJ.

\subsection{AGN core scenario}

VLBI observations of AGN core regions have revealed several aspects that are consistent with the observations of RXJ. AGN typically have a dominating core region, which is compact or marginally resolved, and may also show parsec-scale jets \citep[e.g.,][]{Pushkarev2017_mojave14_shape}. 
Our VLBI observations are consistent with this aspect. However, VLBI cores show optically thick spectra \citep[e.g.,][]{Hovatta2014_mojave11_spectra}, which is inconsistent with the observed negative spectral index of RXJ \citep[see also][]{Yang2022,Goodwin+2025:radioqpe}. This is the strongest observational argument against this scenario.

In addition, other factors disfavor the AGN interpretation as well. While we do not claim a detection of resolved emission, we note that the flux of RXJ is very low ($\sim 200~\mathrm{\mu Jy}$), and typical nearby AGN are brighter and have much higher VLBI brightness temperatures  \citep[][but we note that our brightness temperature is a lower limit]{Hovatta2014_mojave11_spectra}. Furthermore, the astrometry of RXJ shows tentative evidence of a misalignment of the Gaia 1.7\,GHz and 5.0\,GHz emission positions.\footnote{However, see the caveats discussed in \autoref{sec:astrometry}.} 
While the offset between the Gaia and VLBI observations can trivially be explained, for instance by  dust obscuration in the host AGN \citep[e.g.,][]{Plavin2019_gaiavlbi}, the VLBI cores may show a physical offset between frequencies due to the core shift effect \citep[e.g.,][]{Lobanov1998}. 

The core shift effect predicts a frequency-dependent shift of the apparent position on the sky. It is caused by the changing opacity of the expanding jet \citep[e.g.,][]{Blandford1979}, which causes different shells of the jet base to dominate the emission, and thereby shifts the apparent position away from the central black hole. The effect of the core shift is readily detectable by VLBI observations, and is typically on the order of $0.5-1.5~\mathrm{mas}$ for $\sim 1 \mathrm{Jy}$
AGN sources \citep[e.g.,][]{Plavin2019}. 
However, the expected core shift of a source with a Blandford--Koenigl-like jet depends on the flux of the source \citep{Blandford1979, Lobanov1998}:
\begin{equation}
    \Delta R_{\mathrm{mas}} = \dfrac{C_r (1+z)}{D_L\gamma^2 \phi_0} \dfrac{(\nu_2 - \nu1)}{\nu_1\nu_2}\left(\dfrac{L_{\mathrm{syn.}}\sin(\theta)}{\beta(1-\beta \cos(\theta))\Omega}\right)^{2/3}~[\mathrm{mas}].
\end{equation}
Here $C_r = 4.52\times 10^{-12}$ is a constant, $\theta$ is the (unconstrained) viewing angle, $\phi_0 = \phi \csc \theta$ is projected jet opening angle with $\phi\approx 0.5\degree$, and $\Omega=\ln(r_{\mathrm{mas}}/r_c) \approx \ln(100)\approx 5$ \citep[][]{Lobanov1998}. The relation predicts appreciable core shift for radio bright AGN, such as 3C273, 3C216, and 3C345 ($\Delta R_{5-22~\mathrm{GHz}}\approx0.4-0.8~\mathrm{mas}$), which have been confirmed by VLBI studies \citep[e.g.,][]{Lobanov1998, Kovalev2008, Plavin2019}.
While the Doppler factor for a putative core in RXJ is unconstrained, it is clear that RXJ is too faint ($L_{\rm{syn}}\approx 10^{37}~\mathrm{erg/s}$, \citealt{Goodwin+2025:radioqpe}) to have any measurable core shift effect (or would require nonphysically high de-boosting to explain the faintness). 

Finally, RXJ is known as a super-soft X-ray source \citep[e.g.,][]{Giustini+2020:qpe2,Giustini+2024:rxj} suggestive of a radiatively efficient accretion flow, for which canonical jet-like emission would be unusual \citep[e.g.,][]{Sikora2007,Blandford2019}. 

\subsection{Horizon-scale reconnection scenario}
As shown in the previous section, we disfavor emission from a classical AGN jet. \citet{Yang2022} also reached the same conclusions based on the 2017 1.7\,GHz observations, and proposed horizon-scale magnetic reconnection in the accretion disk and episodic jet ejection as source of the VLBI emission, similar to the infrared and X-ray flares seen in Sgr~A* \citep[e.g.,][]{Genzel2003,vonFellenberg2023,vonFellenberg2024}.
Because the emission in this scenario originates from close-to-horizon scales, any observed offset of the emission disfavors the scenario.
However, there are some crucial novelties in this work. First, the astrometric observations in \citet{Yang2022} did not account for the calibrator core shift, which is why the derived astrometric position was consistent with Gaia. Once the calibrator core shift is accounted for, we find an offset between Gaia and 1.7\,GHz VLBI data. In addition, our new higher spatial resolution observations at 5.0\,GHz are also inconsistent with co-spatiality of the radio and optical position. Nevertheless, we point out that there may be multiple reasons for such a Gaia-VLBI offset \citep[see discussion \autoref{sec:astrometry}, and e.g.,][]{Plavin2019_gaiavlbi}. What is more difficult to reconcile is the apparent offset between the 1.7 and 5.0\,GHz emission, which in turn could be caused by astrophysical systematics in the calibrator system (Sect.~\ref{sec:astrometry}). Therefore, we cannot rule out the horizon-scale reconnection scenario observationally. 

\subsection{Remnant outflow or jet scenario}
We propose the following scenario to explain the observed faint emission. The emission results from an optically thin jet component (typically called a jet-knot or jet-blob in VLBI studies) launched from the core of the AGN some time in the recent past, while the core (which we do not observe at present) has extinguished or became otherwise undetectable. This scenario is consistent with most of the observed aspects of the emission \citep[see, e.g., the review by][]{Boccardi2017}:

\begin{itemize}
    \item jet-blobs are typically relatively faint;
    \item jet-blobs are typically optically thin;
    \item jet-blobs are typically compact, but may also be resolved;
    \item jet-blobs may show variability on all timescales; 
    \item jet-blobs typically move at velocities of fractions of milliarcseconds per year.
\end{itemize}

If the Gaia astrometric position is taken as the true position of the SMBH, and the VLBI positions as reliable, the fact that the 5.0 \,GHz is closer to the SMBH than the 1.7\,GHz position is not physical as jet blobs strictly move away from the SMBH (and their emission is optically thin; see, e.g., \autoref{fig:jet-knot}). Therefore, the 1.7\,GHz emission observed in 2017 would have to correspond to a different jet component than the 5.0\,GHz observed in 2023, which is commonly observed in other jet sources \citep[e.g.,][]{Pushkarev2017_mojave14_shape}. Conversely, if the Gaia astrometric position is not taken as the location of the SMBH, which is plausible due to multiple astrophysical biases \citep[e.g.,][]{Plavin2019_gaiavlbi}, the offset between the two different frequencies and observing epochs can be interpreted as motion, and corresponds to $(0.10\pm 0.08)~\mathrm{mas~yr^{-1}}$, or $\sim 0.05~\mathrm{pc~yr^{-1}}\approx0.16~\mathrm{c}$ at $z=0.0237$ \citep{sloan2012}. This projected offset corresponds to scales of a few parsec, or $\sim 10^7~\mathrm{R_{\rm{Schwarzschild}}}$ assuming a black hole mass $\sim 10^6~\mathrm{M_{\odot}}$ \citep[][]{Sun+2013:rxj,Shu2017,Wevers+2022:hosts}.
Due to the unknown viewing angle of the jet and the unknown Doppler amplification of the emission, this is a lower limit on the intrinsic velocity. For reference, this velocity is at the high end of what is typically observed for TDE outflows \citep[e.g.,][]{Cendes+2022:mildlyrel,Cendes+2024:sample,Goodwin+2025:radiotde}, but fully compatible given the high uncertainties. We note that, because of the difficulties disentangling the calibrator core shift in the astrometric measurement, the caveats discussed in \autoref{sec:astrometry} apply.

\section{Conclusions}
In this paper, we presented new VLBI observations of the QPE source RXJ 1301.9+2747 (RXJ). We obtained new high-resolution VLBI data at 5.0\,GHz with the HSA and new VLBA data. In addition, we analyzed archival VLBA data first presented in \cite{Yang2022}.

Our observations show that the radio emission of RXJ is compact and faint. We find tentative evidence for a source motion between 2017 and 2023. While the first two findings are robust, the latter finding strongly depends the assumptions made when calibrating the astrometry, and the astrometric motion should be interpreted with due caution.

We discussed several scenarios for the nature of the radio emission, and disfavor a star formation or AGN-core origin of the emission because the emission is compact and optically thin. We cannot rule out the horizon-scale reconnection scenario proposed by \cite{Yang2022} because the apparent offset between the 1.7 and 5.0\,GHz data may be caused by astrophysical systematics of the calibrator source. Nevertheless, we find this explanation less likely than the remnant outflow or jet scenario, which better aligns with observations indicating that the source is faint, optically thin, compact, and potentially exhibiting motion. 

Our interpretation of a remnant outflow or jet in RXJ is consistent with recent work reported by \citet{Goodwin+2025:radioqpe} that analyzed all available (non-VLBI) radio observations of QPE sources and concluded that they are inconsistent with canonical AGN radio emission, and more similar to remnant TDE-like outflows. We note that in a minority of QPE sources, a precursor optical flare has been observed prior to the discovery of X-ray QPEs \citep{Nicholl+2024:qiz,Chakraborty+2025:upj,Quintin+2023:tormund,Bykov+2024:tormund,Hernandez-Garcia+2025:ansky}, which in the majority of cases can be securely classified spectroscopically as a TDE \citep[with the exception, perhaps, of Ansky;][]{Hernandez-Garcia+2025:ansky}. In the remaining  QPE sources \citep[including the blind X-ray selected,][]{Arcodia+2021:eroqpes,Arcodia+2024:eroqpes,Arcodia+2025:ero5} no precursor optical/UV flares have been found in archival observations. Thus, it is tantalizing that the radio properties of QPE sources may be suggesting that most, if not all of them, are consistent with a remnant jet or outflow (and no strong current activity). This may support the picture that QPEs prefer nuclei with a recent ignition in their nuclei.

\bibliography{bib_rxj}{}
\bibliographystyle{aasjournal}

\appendix
\section{Observations}

\begin{figure}
    \centering
    \includegraphics[width=0.485\textwidth,trim={1cm 1cm 0.5cm 3.5cm},clip]{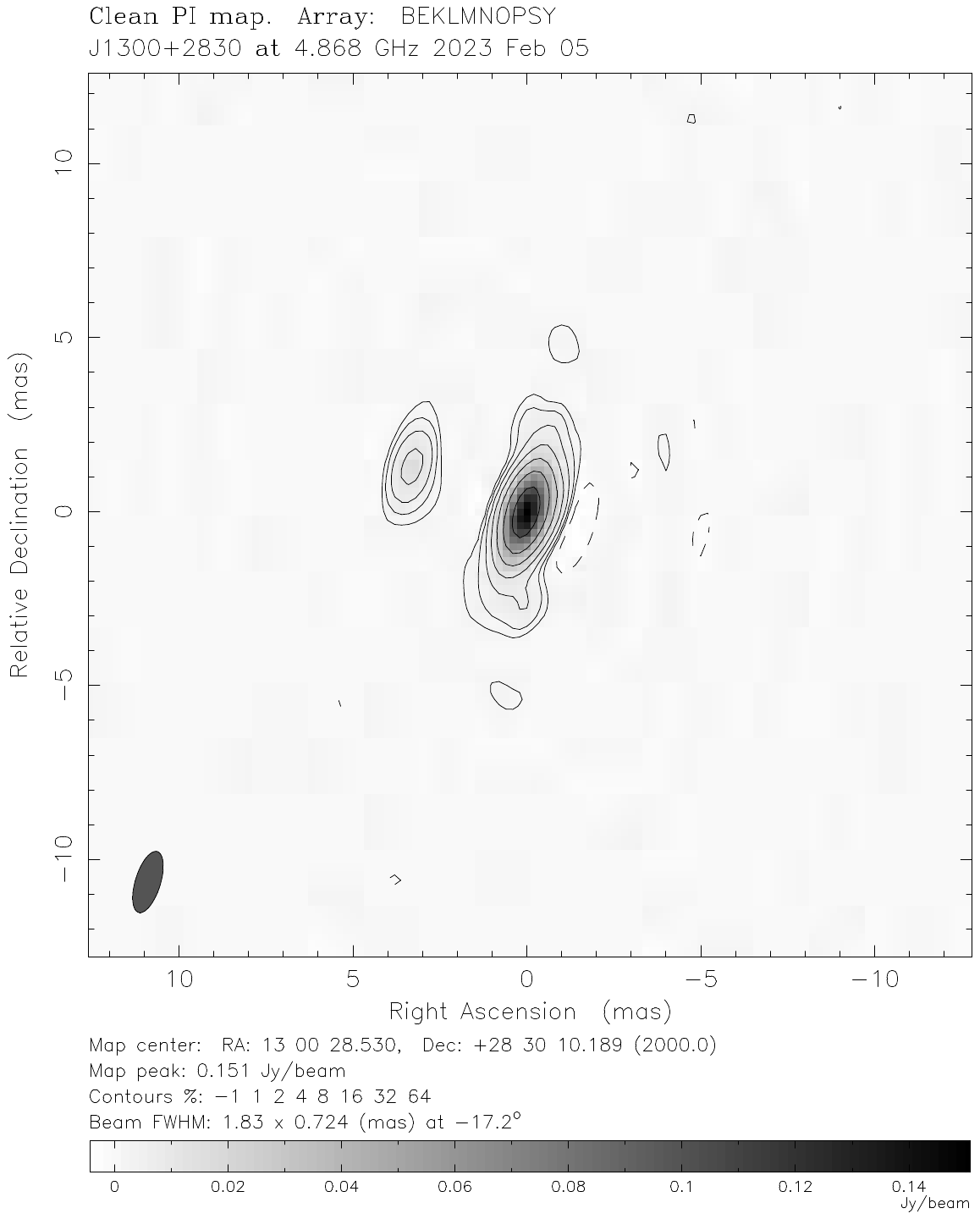}
    \caption{Difmap image of the phase-referencing calibrator J1300+2830.}
    \label{fig:difmap_calib}
\end{figure}

\begin{figure}
    \centering
    \includegraphics[width=0.485\textwidth,trim={1cm 4.2cm 2cm 7cm},clip]{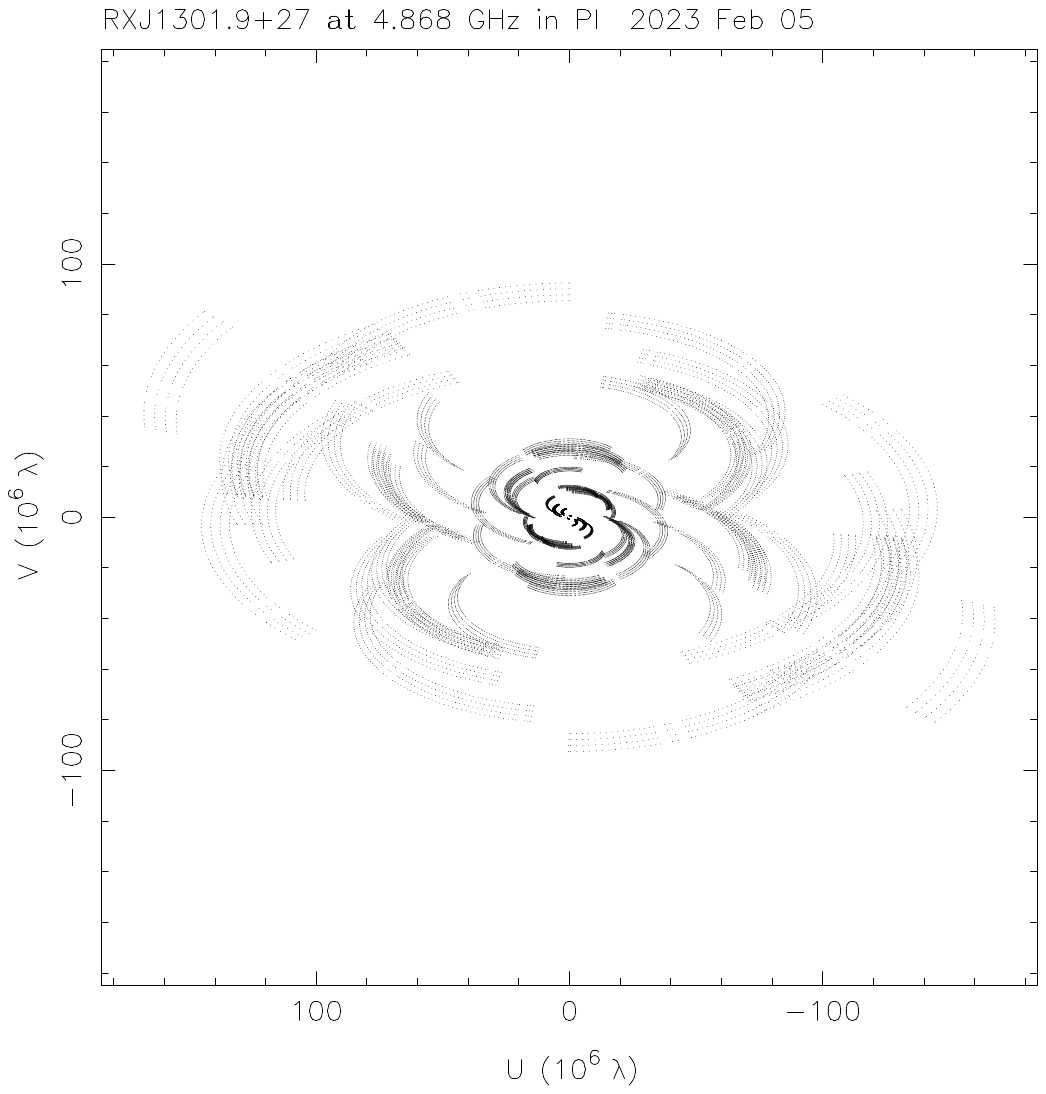}
    \caption{UV coverage of RXJ.}
    \label{fig:uv_coverage}
\end{figure}

\begin{acknowledgements}
This work is part of the M2FINDERS project which has received funding from the European Research Council (ERC) under the European Union’s Horizon 2020 Research and Innovation Programme (grant agreement No 101018682).
S.D.v.F. gratefully acknowledges the support of the Alexander von Humboldt Foundation through a Feodor Lynen  Fellowship and thanks CITA for their hospitality and collaboration.
R.A. was supported by NASA through the NASA Hubble Fellowship grant No. HST-HF2-51499.001-A awarded by the Space Telescope Science Institute, which is operated by the Association of Universities for Research in Astronomy, Incorporated, under NASA contract NAS5-26555.

MG is funded by Spanish MICIU/AEI/10.13039/501100011033 and ERDF/EU grant PID2023-147338NB-C21.

YYK was supported by the MuSES project, which has received funding from the European Union (ERC grant agreement No 101142396). Views and opinions expressed are however those of the author(s) only and do not necessarily reflect those of the European Union or ERCEA. Neither the European Union nor the granting authority can be held responsible for them.

G.M. acknowledges MICIU/AEI/10.13039/501100011033 for support through grants n. PID2020-115325GB-C31 and PID2023-147338NB-C21.

This work has made use of data from the European Space Agency (ESA) mission
{\it Gaia} (\url{https://www.cosmos.esa.int/gaia}), processed by the {\it Gaia} Data Processing and Analysis Consortium (DPAC, \url{https://www.cosmos.esa.int/web/gaia/dpac/consortium}). Funding for the DPAC
has been provided by national institutions, in particular, the institutions
participating in the {\it Gaia} Multilateral Agreement.
LOFAR is the Low Frequency Array designed and constructed by ASTRON. It has observing, data processing, and data storage facilities in several countries, which are owned by various parties (each with their own funding sources), and which are collectively operated by the ILT foundation under a joint scientific policy. The ILT resources have benefited from the following recent major funding sources: CNRS-INSU, Observatoire de Paris and Université d'Orléans, France; BMBF, MIWF-NRW, MPG, Germany; Science Foundation Ireland (SFI), Department of Business, Enterprise and Innovation (DBEI), Ireland; NWO, The Netherlands; The Science and Technology Facilities Council, UK; Ministry of Science and Higher Education, Poland; The Istituto Nazionale di Astrofisica (INAF), Italy.

This research made use of the Dutch national e-infrastructure with support of the SURF Cooperative (e-infra 180169) and the LOFAR e-infra group. The Jülich LOFAR Long Term Archive and the German LOFAR network are both coordinated and operated by the Jülich Supercomputing Centre (JSC), and computing resources on the supercomputer JUWELS at JSC were provided by the Gauss Centre for Supercomputing e.V. (grant CHTB00) through the John von Neumann Institute for Computing (NIC).

This research made use of the University of Hertfordshire high-performance computing facility and the LOFAR-UK computing facility located at the University of Hertfordshire and supported by STFC [ST/P000096/1], and of the Italian LOFAR IT computing infrastructure supported and operated by INAF, and by the Physics Department of Turin university (under an agreement with Consorzio Interuniversitario per la Fisica Spaziale) at the C3S Supercomputing Centre, Italy.
\end{acknowledgements}
\begin{figure}
    \centering
    \includegraphics[width=0.485\textwidth,trim={1cm 1cm 1cm 1.5cm},clip]{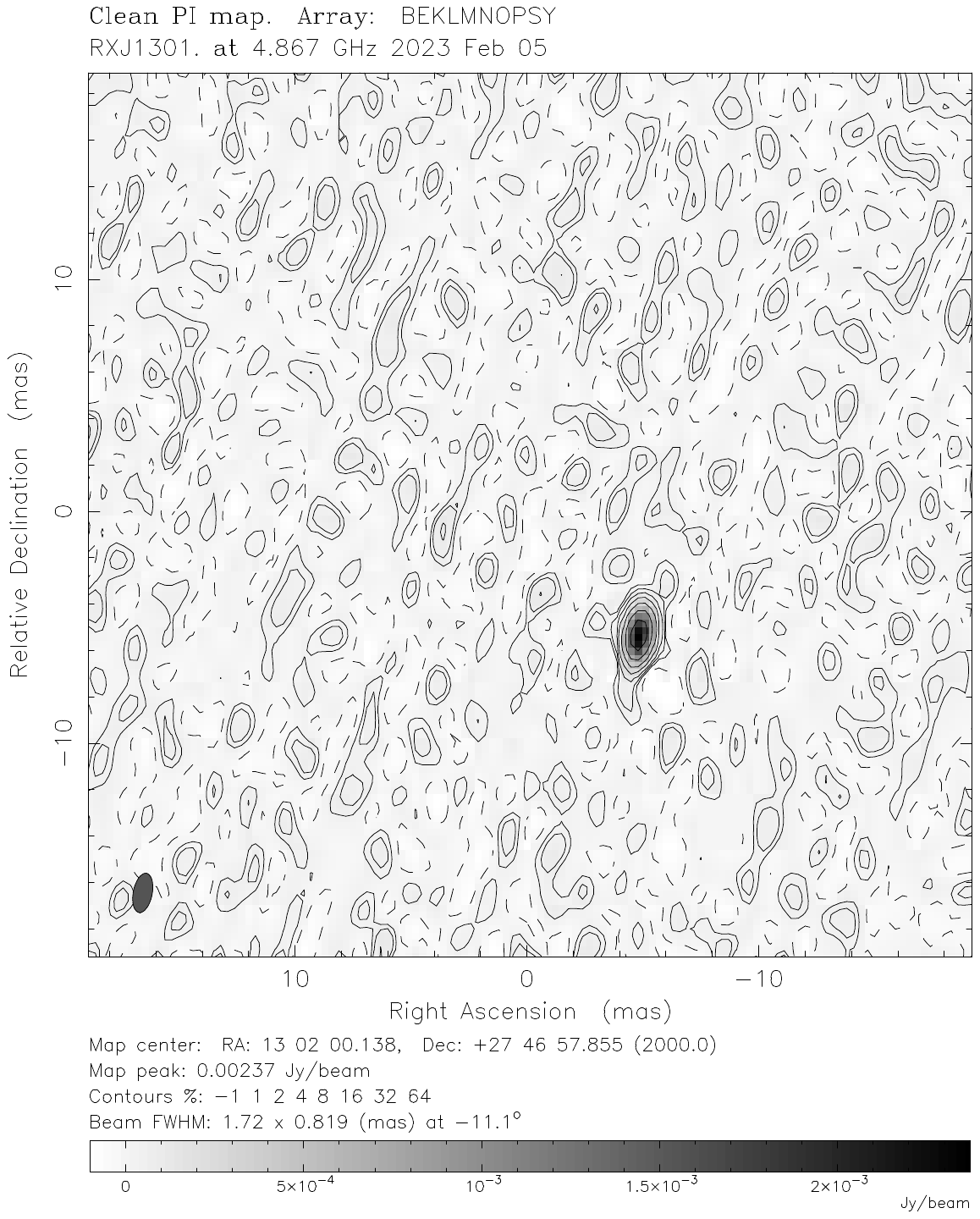}
    \caption{Clean map of RXJ obtained with our new 5.0\,GHz HSA observations. We note that the peak flux derived is likely biased by amplitude self-calibration.}
    \label{fig:clean_map_difmap}
\end{figure}
\autoref{fig:clean_map_difmap} shows the same image as \autoref{fig:clean_map} using the Difmap software. 
\autoref{fig:difmap_calib} shows a clean map of the calibrator, J1300+2830, which shows an extended component in the northeastern direction.
The UV-coverage of RXJ is shown in \autoref{fig:uv_coverage}.

\section{Core shift of the phase reference}\label{app:coreshift}
\begin{figure}
    \centering
    \includegraphics[width=0.485\textwidth]{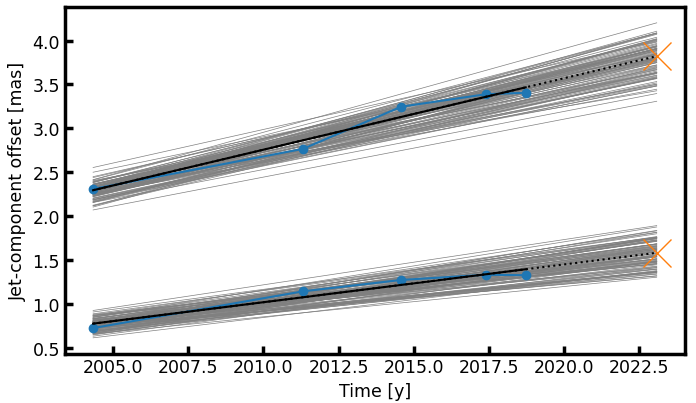}
    \caption{Core to jet-knot separation of the calibrator J1300+2830 as a function of time. The blue points show the R.A. and Dec offset, respectively: the upper one shows R.A., the lower one shows Dec. The gray lines indicate posterior samples of the linear fit.}
    \label{fig:jet-knot}
\end{figure}

\begin{figure}
    \centering
    \includegraphics[width=0.485\textwidth]{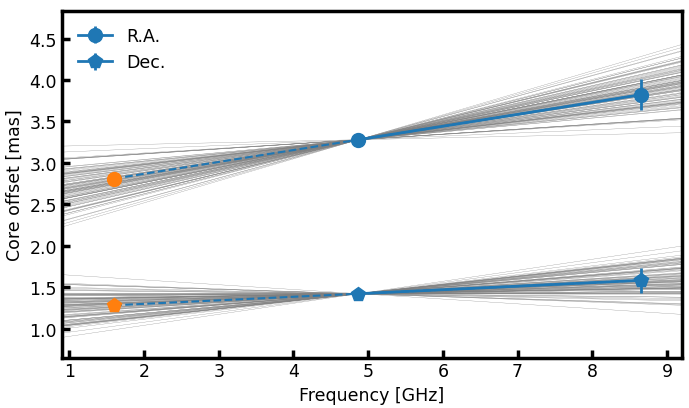}
    \caption{Linear fit to the core to jet component distance at 5.0\,GHz (2023 HSA observations) and the 8.0\,GHz astrogeo measurements scaled to 2023. The orange point shows the extrapolation to 1.7\,GHz. The gray lines indicate posterior samples of linear fit.}
    \label{fig:core-shift-calibrator}
\end{figure}

We show the motion of the jet component of the calibrator in \autoref{fig:jet-knot}. The motion was derived by measuring the core to jet component distance as function of time in the calibrated fits files provided on the astrogeo VLBI calibrator archive observed at 8\,GHz. The linear motion of the jet component is extrapolated to 2023, the date of the HSA 5.0\,GHz observations.
\autoref{fig:core-shift-calibrator} shows the 8.0\,GHz core to jet component distance, and a linear extrapolation between the 5.0\,GHz and 8.0\,GHz to 1.7\,GHz, the frequency of the 2017 VLBI observations of RXJ.
We can test this extrapolation using an additional L-band observation VLBA (BY182) observation of the calibrator. This observation was taken close in time (12 December, 2023). The observation was short in duration, and the spatial resolution is thus limited $\sigma_{\rm{beam}}=(10.0\times3.3)~\mathrm{mas}^2$.
With this resolution, the jet component is only marginally resolved. Even so, when cleaning and self-calibrating the data we find a clean component offset by $(2.5, 1.2)~\mathrm{mas}$ from the core, in good agreement with the extrapolated values derived from the astrogeo observations. The nominal astrometric uncertainty is $\pm0.1~\mathrm{mas}$, but it is likely underestimated because the component is confused with the core. Still, the broad agreement between the two different methods indicates that the extrapolation is valid.

\end{document}